\documentclass[
twocolumn,
showpacs,
superscriptaddress,
nofootinbib,
amssymb,
a4paper
]{revtex4-1}
\usepackage[dvipdfmx]{graphicx}
\usepackage{ulem}
\usepackage{color}
\usepackage{bm}
\usepackage{braket}
\usepackage{amsmath,amssymb}
\usepackage{comment}
\usepackage{lineno}

\def\({\left(}
\def\){\right)} 
\def\[{\left[}
\def\]{\right]} 
\newcommand{\mcal}[1]{{\mathcal{ #1 }}}
\newcommand{\slh}[1]{ {#1} \hspace{-0.65em}/} 
\newcommand{\Tr}{\mathrm{Tr}}

\newcommand{\Det}{\mathrm{Det}}
\newcommand{\hy}{\mathchar`-} 
\newcommand{\g}{\gamma} 

\newcommand{\inst}{{\rm I}}
\newcommand{\ainst}{{\rm \bar{I}}}
\newcommand{\UA}{${\rm U}(1)_A$~}

\begin{document}
%
%
%
\title{
  Possible scenario of dynamical chiral symmetry breaking in the interacting instanton liquid model
}
%
\author{
Yamato Suda
}
\affiliation{
  Department of Physics, 
  Tokyo Institute of Technology, 
  2-12-1 Ookayama, Megro, Tokyo 152-8551, 
  Japan
}
\author{
Daisuke Jido
}
\affiliation{
  Department of Physics, 
  Tokyo Institute of Technology, 
  2-12-1 Ookayama, Megro, Tokyo 152-8551, 
  Japan
}
%
%
\date{\today}

\begin{abstract}
  We compute the vacuum energy density as a 
function of the quark condensate in the interacting 
instanton liquid model (IILM) and examine the pattern 
of dynamical chiral symmetry breaking from its 
behavior around the origin. 
This evaluation is performed by using simulation results of the IILM. 
We find that chiral symmetry is broken in 
the \UA anomaly assisted way in the IILM with three-flavor dynamical quarks. 
We call such a symmetry breaking the anomaly-driven breaking 
which is one of the scenarios 
of chiral symmetry breaking proposed in the context of 
the chiral effective theories.
We also find that the instanton-quark interaction 
included in the IILM plays a crucial role for 
the anomaly-driven breaking by comparing 
the full and the quenched IILM calculations.

\end{abstract}
%
%
\maketitle
%
%

\section  {Introduction}
\label{sec:Introduction}

  In hadron physics, 
chiral symmetry and its dynamical breaking have 
led to our systematic understanding of hadron spectra.
For instance, 
the arrangement of light pseudoscalar $(\pi,K,\bar{K},\eta)$-mesons 
can be explained through the dynamical breaking of 
${\rm SU}(3)_L \times {\rm SU}(3)_R$ flavor symmetry
as the Nambu-Goldstone theorem states.
On the other hand, mesons like $f_0(500)$ are still 
puzzles in hadron physics \cite{Pelaez:2016,PDG:2022}.
The characteristics of $f_0(500)$, such as its existence,
mass and internal structure, have been under discussion 
for a long time \cite{Jaffe:1977,Oller:1999,Black:1999,Ishida:1999,Colangelo:2001,Kunihiro:2004,Napsuciale:2004,Pelaez:2004,Mennessier:2011,Caprini:2006,Pennington:2007,Fariborz:2009,Hyodo:2010,Mennessier:2011,Parganlija:2013,Pelaez:2017,Ablikim:2019,Soni:2020,Achasov:2021,Pelaez:2021}. 
The $\sigma$ meson that is recognized 
as the chiral partner of the pion is one of 
the candidates of $f_0(500)$ because it should exist in 
the state $J^P=0^+$ if chiral symmetry is manifested 
in the low-energy regime of quantum chromodynamics (QCD).
Since the $\sigma$ meson emerges as a fluctuation 
of an order parameter of chiral symmetry, 
there is a growing need to systematically 
investigate the chiral dynamics.

  In revealing such hadron properties, we need to 
understand the pattern of dynamical chiral symmetry 
breaking of QCD.
Many models have been proposed to describe 
the dynamical chiral symmetry breaking
\cite{GellMann:1960,Nambu:1961,Verbaarschot:2000}.
A notable example is the Nambu-Jona-Lasinio (NJL) model which 
exhibits essential features 
of QCD and provides a mathematically traceable framework 
under the mean field approximation.
In this model, the attractive interaction among quark fields 
induces the quark condensate in the vacuum and it breaks chiral symmetry. 
As chiral symmetry is broken dynamically, 
pseudoscalar mesons become the massless 
Nambu-Goldstone bosons, while their chiral partners, 
the scalar mesons, acquire finite masses. 
Furthermore, the inclusion of the \UA anomaly 
enables us to comprehend the spectrum of 
pseudoscalar mesons including $\eta$ and $\eta'$ 
\cite{Witten:1979,tHooft:1976_a,tHooft:1976_b}.
Comprehensive understanding of the pattern
and mechanism of dynamical chiral symmetry breaking 
is one of the keys to unraveling the properties 
of hadrons and the QCD vacuum on which they rely.

  Recently, the chiral effective theories 
have shown that the different patterns of 
dynamical chiral symmetry breaking predict 
the different values of the $\sigma$ mass \cite{Kono:2021}.
In the NJL model without the \UA anomaly term,
chiral symmetry breaks dynamically when the interquark attractive 
interaction is sufficiently strong. With the presence of 
the \UA anomaly term, chiral symmetry is broken dynamically 
even if the four-point interquark interaction is smaller than its 
critical value as long as the contribution from 
the \UA anomaly is significant 
(which we call anomaly-driven symmetry breaking). 
The mass of the $\sigma$ meson
assumed as the chiral partner of the pion shows 
that in the former case it is larger than approximately $800~{\rm MeV}$, 
while in the latter case it is smaller than 
about $800~{\rm MeV}$~\cite{Kono:2021}. 
These findings are derived 
from analyses using the linear sigma model and 
the NJL model, while the universality of 
their consequence remains unclear in other systems.

  In this paper, we examine whether 
the anomaly-driven breaking of chiral symmetry universally realizes 
in the context of other than the chiral effective models.
The realization of the anomaly-driven breaking 
is identified by the quark condensate 
dependence of the vacuum energy density.
This characteristic is deduced from the discussion
within the chiral effective models~\cite{Kono:2021}. 
We generalize this concept to validate it 
in other models using the curvature of 
the vacuum energy density with respect 
to the quark condensate at the origin.
As an example of its application,
we use the interacting instanton liquid model (IILM).
The IILM is a phenomenological model 
that describes the chiral symmetry breaking 
by treating the QCD vacuum as a superposition of instantons \cite{Shuryak:1982_1,Shuryak:1982_2}.
Since the instanton results in the \UA anomaly~\cite{tHooft:1986}, 
we discuss the possibility of the anomaly-driven symmetry breaking 
using the IILM that has both features like the chiral symmetry breaking 
and the \UA anomaly related to the instantons.

  This paper is organized as follows.
In Sect.~\ref{sec:AnomalyDrivenBreaking}, 
we introduce the definition of anomaly-driven 
breaking of chiral symmetry in the NJL model and 
other models.
In Sect.~\ref{sec:formulation}, we describe 
the formulation of the IILM that is used in this study 
and details the numerical simulations.
In Sect.~\ref{sec:Results}, we present the simulation 
results of the IILM and discuss their interpretations.
In Sect.~\ref{sec:Conclusions}, we conclude this paper.

\section  {Anomaly-driven breaking of chiral symmetry}
\label{sec:AnomalyDrivenBreaking}

  In this section, we introduce the definition of 
the anomaly-driven chiral symmetry breaking in the NJL model 
and other models.

\subsection  {Anomaly-driven breaking in the NJL model}
\label{subsec:ADBinNJL}

  We briefly explain the anomaly-driven breaking 
of chiral symmetry which was proposed by 
the previous work \cite{Kono:2021}
based on the three flavor NJL model with the axial anomaly term. 
In Ref.~\cite{Kono:2021}, the following Lagrangian is considered:
\begin{eqnarray}
  \mcal{L}_{\rm NJL} &=& 
  \bar{\psi} 
  (i\gamma^{\mu} \partial_{\mu} - \mcal{M})
  \psi \nonumber \\ 
  &\ & + 
  g_S \sum_{a=0}^8
  \[
  \(\bar{\psi}\frac{\lambda_a}{2}\psi\)^2
  +
  \(\bar{\psi}i \gamma_5 \frac{\lambda_a}{2} \psi\)^2
  \] \nonumber \\ 
  &\ & +
  \frac{g_D}{2} 
  \left\{
  \det_{i,j}\[ \bar{\psi}_i (1-\gamma_5) \psi_j \]
  +
  {\rm H.c.}
  \right\}, \label{eq:NJL_lagrangian}
\end{eqnarray}
for the quark fields $\psi = (u, d, s)^T$, 
where $\mcal{M}$ is the quark mass matrix given 
by $\mcal{M}={\rm diag}(m_q,m_q,m_s)$ with assuming 
isospin symmetry $m_q=m_u=m_d$, $\lambda_a\ (a=0,1,\dots,8)$ 
represent the Gell-Mann matrices 
in the flavor space with $\lambda_0=\sqrt{2/3}\bm{1}$, 
$\det_{i,j}$ is understood as determinant operation 
over the flavor indices of the quark fields 
$\psi_i, \bar{\psi}_j\ (i,j=u,d,s)$,
and $g_S,g_D$ are the coupling constants 
for the four-point vertex interaction and the 
determinant-type \UA breaking interaction, 
respectively. 
In the mean field approximation, 
the effective potential is obtained from 
the Lagrangian (\ref{eq:NJL_lagrangian}) 
as a function of the dynamical quark mass $M$
as follows 
\begin{eqnarray}
  &\ & V_{\rm eff}(M) \nonumber \\
  &=& i N_c
  \sum_{f=q,q,s} \int \frac{d^4p}{(2\pi)^4} \Tr 
  \[
  \ln (\g \cdot p -M_f)
  +
  \frac{\g \cdot p -m_f}{\g \cdot p - M_f}
  \] \nonumber \\
  &\ &
  - \frac{g_S}{2} (2\braket{\bar{q}q}^2 + \braket{\bar{s}s})
  - g_D \braket{\bar{q}q}^2\braket{\bar{s}s}, \label{eq:EffV_def}
\end{eqnarray}
where $\braket{\bar{q}q} = \braket{\bar{u}u} = \braket{\bar{d}d}$ 
is the quark condensate with isospin symmetry and is given 
as a function of $M$. Their specific forms are given 
by Eq.~(A.1) and Eq.~(A.2) in Ref.~\cite{Kono:2021}.

  Following Ref.~\cite{Kono:2021}, we start with 
the case of the chiral limit and no anomaly term, i.e., 
$m_q=m_s=0$ and $g_D=0$ in Eq.~(\ref{eq:NJL_lagrangian}).  
In this situation, chiral symmetry is dynamically broken 
at the vacuum with a finite dynamical quark mass 
when the dimensionless coupling constant $G_S\equiv g_S/g_S^{\rm crit}$ 
is larger than $1$. Here, $g_S^{\rm crit}$ is 
the critical coupling constant defined 
by $g_S^{\rm crit} = 2\pi^2/(3\Lambda_3^2)$ 
with a three-momentum cutoff $\Lambda_3$ 
for the quark loop function.
In other words, if there exists sufficiently 
strong four-quark interaction, chiral symmetry 
is dynamically broken in the vacuum.
This is the well-known scenario of the dynamical 
chiral symmetry breaking in the NJL model.

  Next, let us take the anomaly term into account 
with $g_D \neq 0$ in Eq.~(\ref{eq:NJL_lagrangian}) 
while keeping the chiral limit.
In this situation, even though $G_S$ is less than $1$,
chiral symmetry can be broken dynamically in the vacuum
due to the existence of the axial anomaly term.
The previous work \cite{Kono:2021} demonstrated 
that such a situation is realized 
with a sufficiently large contribution 
from the anomaly term.
We refer to such chiral symmetry breaking 
as the anomaly-driven breaking of chiral symmetry or 
shortly the anomaly-driven breaking in this paper.

  The above patterns of the chiral symmetry breaking
are related to the hadronic properties, 
such as the mass of the $\sigma$ meson \cite{Kono:2021}.
In the literature, in order to study the relationship more quantitatively, 
the explicit chiral symmetry breaking by finite current quark mass is introduced 
and the $\sigma$ meson is assumed to be the chiral partner of the pion.
The authors calculated the $\sigma$ meson mass with varying values of 
the dimensionless couplings $G_S$ 
and $G_D\equiv \Lambda_3 g_D/(g_S^{\rm crit})^2$ 
so as to reproduce the $\eta'$ mass. 
As a result, they found that the $\sigma$ mass is smaller than 
about $800~{\rm MeV}$ when the anomaly-driven breaking is realized, 
i.e., $G_S<1$, and larger than about $800~{\rm MeV}$ if the normal 
breaking is done, i.e., $G_S>1$. 

  According to the previous work \cite{Kono:2021}, 
the definition of the anomaly-driven breaking 
in the NJL model is that the chiral symmetry is 
dynamically broken even though 
the dimensionless coupling constant $G_S$ is less than $1$.
However, the definition of this determination procedure 
relies on the model-specific parameter $G_S$, 
which makes its application to other models nontrivial.
To discuss the anomaly-driven breaking 
in other models and systems, in the next section, 
we generalize the definition of 
the determination procedure 
for the anomaly-driven breaking.

\subsection  {The anomaly-driven breaking in other models}
\label{subsec:ADbreaking_general}

  In this subsection, we generalize the definition of 
the determination procedure for anomaly-driven breaking 
based on arguments in the NJL model. 
The anomaly-driven breaking of chiral symmetry 
was initially introduced in the chiral effective models using 
the model-specific coupling constants. 
Here, we present a key quantity that links the model-specific 
and the model-independent determination procedures 
for the anomaly-driven breaking. 
That is the sign of the curvature of the effective potential 
at the point with zero quark condensate. 
We use the latter as the definition of determination 
procedure for the anomaly-driven breaking in this paper.

  We first consider the NJL model with no anomaly term 
in the vanishing quark mass limit $(m \to 0)$.
In this case, as we have mentioned in the last section, 
chiral symmetry is dynamically broken in the vacuum 
when a dimensionless four-quark coupling $G_S$ is 
greater than $1$.
Here, the evaluation of the second derivative of 
the effective potential (\ref{eq:EffV_def}) 
with respect to the quark condensate at 
the point $\braket{\bar{q}q}=0$ yields 
the following result:
\begin{eqnarray}
  \left. \frac{\partial^2 V_{\rm eff}}{\partial \braket{\bar{q}q}^2} \right|_{\braket{\bar{q}q}=0} 
  = g_S^{\rm crit} - g_S = g_S^{\rm crit}\(1-G_S\),
  \label{eq:2nd_deriv_V}
\end{eqnarray}
where we use $G_S=g_S/g_S^{\rm crit}$.
From Eq.~(\ref{eq:2nd_deriv_V}), we find that 
the parameter region where chiral symmetry 
is dynamically broken is linked to the negativity 
of the curvature of the effective potential at the origin.
We refer to such chiral symmetry breaking as the normal breaking 
in contrast to the anomaly-driven one.

  Next, let us turn on the anomaly term 
by $g_D \neq 0$ while keeping the chiral limit. 
Again, chiral symmetry can be dynamically broken and that 
leads to the nonzero quark condensate in the vacuum. 
The main difference compared to the case without 
the anomaly term is that chiral symmetry can be 
broken dynamically even though the dimensionless coupling $G_S$ 
is less than $1$. 
Calculating the second derivative of 
the effective potential (\ref{eq:EffV_def})
with the anomaly term, we obtain 
the same result to Eq.~(\ref{eq:2nd_deriv_V}). 
Thus, the pattern of dynamical chiral symmetry 
breaking which is distinguished whether $G_S$ 
is greater than $1$ or not 
corresponds directly to the sign of 
the curvature (\ref{eq:2nd_deriv_V}).

\begin{figure}
  \includegraphics[width=86mm]{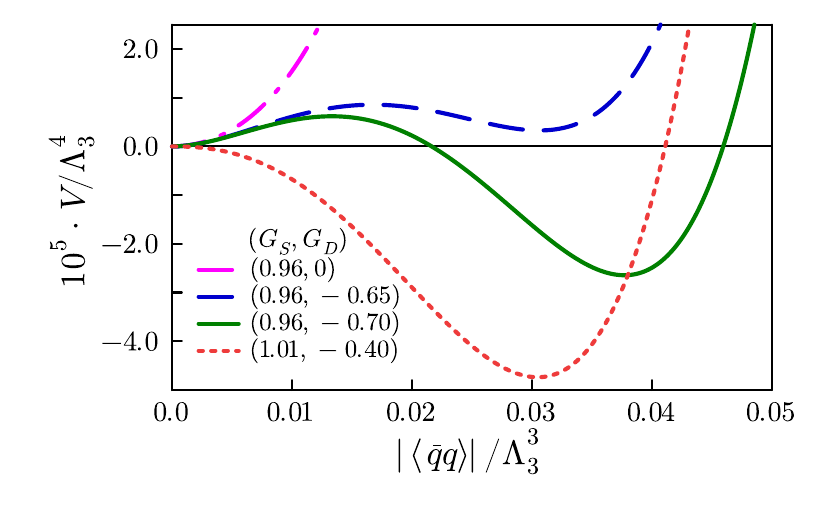}
  \caption{
    \label{fig:GSGD_validation} 
    The effective potentials as a function of
    the quark condensate normalized by 
    the cutoff scale $\Lambda_3^3$.
    The mean field approximation 
    is used and the chiral limit is assumed.
    For the effective potential, 
    the irrelevant constant is subtracted 
    from that. 
    The different coupling 
    constants are plotted, 
    $(G_S,G_D)=(0.96,0.0)$ [dashed-dotted, magenta];
    $(0.96,-0.65)$ [dashed, blue]; 
    $(0.96,-0.70)$ [solid, green]; 
    $(1.01,-0.40)$ [dotted, red].
    }
\end{figure}

  In Fig.~\ref{fig:GSGD_validation}, we show the 
effective potentials as a function of the absolute value of 
quark condensate for four parameter sets in the chiral limit. 
In order to set the origin to zero, irrelevant constants are 
subtracted from the effective potentials.
We can see that when the dimensionless coupling $G_S$ 
is greater than $1$, the curvature of effective potential 
at the origin is negative (red dotted line). 
On the other hand, when chiral symmetry is dynamically broken 
even though $G_S$ is less than $1$ (green solid line), 
the curvature is positive at the origin. For the remaining 
two parameter sets (magenta dash-dotted and blue dashed line), 
the effective potential has a minimum value at the origin and thus 
chiral symmetry is not broken in the vacuum.
In this way, when the chiral symmetry is dynamically broken, 
the dimensionless coupling constant $G_S$ 
and the curvature of the effective potential 
at the origin are linked in the NJL model 
including the anomaly term for the chiral limit.

  We also confirm that such a relationship between 
the dimensionless coupling $G_S$ and the curvature of the 
effective potential remains unchanged even if 
small current quark masses are introduced.
The current quark mass dependence appears 
in the effective potential as a product with the 
linear term of the quark condensate for small quark masses. 
That term vanishes by performing the second derivative 
with respect to the quark condensate. 
Equation~(\ref{eq:2nd_deriv_V}) thus remains the same 
in the presence of small current quark masses.

  Based on the arguments in the NJL model, 
we use the curvature of the effective potential 
to determine whether the anomaly-driven breaking or not, 
instead of the model specific coupling constant.
Since the effective potential is proportional to 
the vacuum energy density $\epsilon$ except for 
a constant in field theory, 
we take the inequality 
\begin{eqnarray}
  \left. \frac{\partial^2 \epsilon}{\partial \braket{\bar{q}q}^2} \right|_{\braket{\bar{q}q}=0} > 0,
  \label{eq:Condition_ADB}
\end{eqnarray}
to be the definition of 
the determination procedure 
for the anomaly-driven breaking 
also for finite quark masses.
We apply this definition to 
the instanton liquid model 
and discuss the feasibility 
of the anomaly-driven 
chiral symmetry breaking.

\section  {Formulation}
\label{sec:formulation}

  In the present section, we describe the model 
and numerical calculation method we used.

\subsection  {Interacting instanton liquid model}
\label{subsec:IILM}

  The QCD Euclid partition function 
is approximated by the superposition 
of instantons \cite{Schafer:1996}. 
The canonical partition function is given as
\begin{eqnarray}
  Z &=& \frac{1}{N_+! N_-!} 
  \int \prod_{i=1}^{N_+ + N_-} \[ d\Omega_i f(\rho_i) \]
  \exp (-S_{\rm int}) \nonumber \\ 
  &\ & \times \prod_{f=1}^{N_f} 
  \Det (\slh{D} + m_f).
  \label{eq:IILM}
\end{eqnarray}
Here, $N_+$ and $N_-$ are the numbers of instantons and 
anti-instantons, respectively, $d\Omega_i=dU_id \rho_i d^4z_i$ 
is the measure of the path integral over the 
collective coordinate space for the color orientation 
in the color ${\rm SU}(N_c)$ group, size and position 
associated with the $i$-th instanton, 
and $f(\rho)$ is semiclassical instanton amplitude. 
The interactions between instanton-instanton 
and instanton-quark are included 
in $S_{\rm int}$ and $\Det(\slh{D} + m_f)$, respectively.
$D_\mu$ and $m_f$ represent the covariant derivative 
for the quark fields and the current quark mass of flavor $f$, 
respectively.

  The explicit expression of the semiclassical 
instanton amplitude calculated by 't Hooft~\cite{tHooft:1976_b}
reads
\begin{eqnarray}
  f(\rho) 
  &=& C_{N_c} \[ \frac{8\pi^2}{g^2(\rho)} \]^{2N_c} 
  \exp \[- \frac{8\pi^2}{g^2(\rho)}\]
  \frac{1}{\rho^5} \nonumber \\
  &=& C_{N_c} \frac{1}{\rho^5} 
  \beta_1(\rho)^{2N_c} \nonumber \\
  &\ & \times 
  \exp \[ 
  - \beta_2(\rho) 
  + \( 2N_c - \frac{b'}{2b} \) \frac{b'}{2b} 
  \frac{1}{\beta_1(\rho)} \ln \beta_1(\rho)
  \] \nonumber \\ 
  &\ & + \mcal{O}(3 \hy {\rm loop}), \label{eq:inst_density} \\
  C_{N_c} &=&  \frac{0.466 e^{-1.679N_c}}{(N_c-1)!(N_c-2)!}, 
  \label{eq:CNc}
\end{eqnarray}
where the gauge coupling $g^2$ is given 
as a function of the instanton size $\rho$, 
the beta functions $\beta_1,\beta_2$ include 
up to 2-loop order and the Gell-Mann--Low 
coefficients are given as follows:
\begin{eqnarray}
  \beta_1(\rho) &=& -b \ln (\rho \Lambda),\hspace{1em} 
  \beta_2(\rho) = \beta_1(\rho) 
  + \frac{b'}{2b} \ln \[\frac{2}{b} \beta_1(\rho)\], 
  \label{eq:1loop} \nonumber \\ \\
  b &=& \frac{11}{3}N_c- \frac{2}{3}N_f, \hspace{1em}
  b' = \frac{34}{3} N_c^2 - \frac{13}{3} N_c N_f 
  + \frac{N_f}{N_c}. \label{eq:2loop} \nonumber \\
\end{eqnarray}
Here, $N_c$ and $N_f$ denote 
the number of colors and 
of the dynamical quark flavors, 
respectively, and $\Lambda$ is the scale 
parameter which is introduced for calculating $\beta$-function  
by using the Pauli-Villars regularization scheme \cite{Schafer:1996}.

  The instanton-instanton interaction $S_{\rm int}$ 
in Eq.~(\ref{eq:IILM}) is expressed 
by the sum of all possible pairs of instantons
\begin{eqnarray}
  S_{\rm int} 
  = \sum_{l<m, l \neq m}^{N_+ + N_-} 
  S^{(2)}_{\rm int}(l,m), 
  \label{eq:Sint}
\end{eqnarray}
where $l,m$ refer to each instanton or anti-instanton 
in the ensemble. The two-body interaction 
between instantons (or anti-instantons) 
$S^{(2)}_{\rm int}(l,m)$ is defined as the 
difference between the action of 
two-instanton configuration 
and twice of the single instanton action $S_0(\rho)=8 \pi / g^2(\rho)$:
\begin{eqnarray}
  S^{(2)}_{\rm int}(l,m) = S[A_\mu(l,m)] - 2S_0. 
  \label{eq:Twobody_int}
\end{eqnarray}
Two-instanton configuration 
is no longer an exact solution to the 
classical Yang-Mills equations.
To calculate the two-body interaction, 
one uses an {\it Ansatz} for the 
gauge configurations.
This idea was developed by Sch\"afer and Shuryak \cite{Schafer:1996,Schafer:1998}.
We use the streamline {\it Ansatz} \cite{Yung:1988,Verbaarschot:1991} 
and its form of $S_{\rm int}$ is given by Eq.~(A5) in Ref.~\cite{Schafer:1996}.

  The instanton-quark interaction 
represented by the determinant in Eq.~(\ref{eq:IILM}) 
is evaluated by factorizing it into two parts,
a high and a low momentum parts as
\begin{eqnarray}
  \Det (\slh{D} + m_f) 
  &=& \Det_{\rm high}(\slh{D} + m_f) 
  \Det_{\rm low}(\slh{D} + m_f).~~~~
\end{eqnarray}
The high momentum part is evaluated as 
the product of contributions of each instanton 
calculated by using the Gaussian approximation. 
The low momentum part is evaluated by using 
the quark zero-mode wave functions 
in the instanton and anti-instanton backgrounds \cite{Schafer:1996}. 
As a result, the instanton-quark interaction 
takes the following form:
\begin{eqnarray}
  &\ & \Det (\slh{D} + m_f) 
  = \(\prod_{i=1}^{N_+ + N_-} 1.34 \rho_i\) 
  \Det_{\inst, \ainst} (-iT + m_f \bm{1}), \nonumber \\ 
  \label{eq:Det}
  &\ & T = 
  \begin{pmatrix}
  \bm{0}_{N_+ \times N_+}       & (\mcal{T})_{N_+ \times N_-} \vspace{1em}\\ 
  (\mcal{T}^\dagger)_{N_- \times N_+}  & \bm{0}_{N_- \times N_-}  \\ 
  \end{pmatrix}, \label{eq:matrix_T} \\
  &\ & (\mcal{T})_{IJ} 
  = \int d^4x \, \psi^*_{0,I}(x;U,\rho,z) i \gamma_{\mu} D_\mu \psi_{0,J}(x;U,\rho,z).~~~~
\end{eqnarray}
Here $T$ is called the overlap matrix with a size of  
$(N_+ + N_-) \times (N_+ + N_-)$.
This matrix is spanned 
by the quark zero-mode wave functions $\psi_{0,I}(x;U,\rho,z)$
in the instanton ($\inst$) and anti-instanton $(\ainst)$ backgrounds
which are labeled by the collective coordinate $\{U,\rho,z\}$.
The specific forms of the quark zero-mode wavefunctions 
are summarized in Appendix~2 of Ref.~\cite{Schafer:1998}. 
The determinant operation $\Det_{\inst, \ainst}$ is 
taken over the space represented in 
Eq.~(\ref{eq:matrix_T}) and $m_f \bm{1}$ is 
understood as a diagonal matrix of 
$(N_+ + N_-) \times (N_+ + N_-)$.
The explicit expression of $\mcal{T}$ is 
given by Eq.~(B2) in Ref.~\cite{Schafer:1996}.

  In this paper, we work on the ${\rm SU(3)}$ flavor 
symmetric limit, i.e., the number of quark flavors 
is set to $N_f=3$ and the current quark masses 
are equally set to $m=m_f~(f=1,2,3)$. 
The quark condensate $\braket{\bar{q}q}$ represents
the one-flavor amount for our calculations.

\subsection  {Free energy in instanton ensemble}
\label{subsec:F}

  The vacuum energy density is identified 
the free energy density as $\epsilon = F$ 
at zero temperature and we simply call it 
the free energy denoted as $F$ in what follows.
This relationship is derived from 
the standard thermodynamics relation. 
The free energy is expressed as 
\begin{eqnarray}
  F = - \frac{1}{V} \ln Z, \label{eq:F_def}
\end{eqnarray}
with the four-dimensional space-time volume $V=L^4$ 
and the partition function $Z$ of the system considered.

  We explain the method to compute 
the value of the partition function $Z$. 
The method that we use is well-known as the thermodynamics 
integration method, and it has been applied to 
the IILM calculation in the previous work \cite{Schafer:1996}.
In this method, one writes the effective action as 
\begin{eqnarray}
  S_{\rm eff}(\alpha) = S_{\rm eff}(0) + \alpha S_1, \label{eq:Seff_decomp}
\end{eqnarray}
with the partition function 
$Z(\alpha) = \int d\Omega \exp\[-S_{\rm eff}(\alpha)\]$. 
The desired partition function $Z$ is reproduced as $Z(\alpha=1)$.
This form of the effective action (\ref{eq:Seff_decomp}) 
interpolates between a known solvable action $S_{\rm eff}(0)\equiv S_{\rm eff}^0$ 
and the full action $S_{\rm eff}(1)=S_{\rm eff}^0+S_1$. 
One obtains the partition function $Z(\alpha=1)$
straightforwardly as
\begin{eqnarray}
  \ln \[ Z(\alpha=1)\] 
  = \ln \[Z(\alpha=0)\] - \int_0^1 d\alpha \braket{0|S_1|0}_{\alpha},~~
  \label{eq:Z_alpha_cal}
\end{eqnarray}
where the expectation value $\braket{0|O|0}_\alpha$ 
is defined by 
\begin{eqnarray}
  \braket{0|O|0}_\alpha \equiv 
  \frac{1}{Z(\alpha)} \int d \Omega~ O(\Omega) e^{-S_{\rm eff}(\alpha)},
  \label{eq:alpha_EV_def}
\end{eqnarray}
with configurations according 
to $p(\Omega_i) \propto \exp\[-S_{\rm eff}(\alpha)\]$. 
Therefore, if we know the decomposition (\ref{eq:Seff_decomp}) 
and the values of $S_{\rm eff}^0$ and $Z(\alpha=0)$, 
we can compute the partition function $Z(\alpha=1)$ from 
Eq.~(\ref{eq:Z_alpha_cal}). 

  For our computation of the partition function 
in the instanton liquid model, 
we use the following decomposition 
of the effective action $S_{\rm eff}$ 
as Ref.~\cite{Schafer:1996}:
\begin{eqnarray}
  S_{\rm eff} (\alpha)
  &=& \sum_{i=1}^{N_+ + N_-} \left\{ \ln \[f (\rho_i)\] 
  + (1-\alpha) \nu \frac{\rho_i^2}{\bar{\rho}^2} \right\} \nonumber \\
  &\ & +\alpha\[ S_{\rm int} - \sum_{f=1}^{N_f} \ln \Det ( \slh{D} +m_f) \],
  \label{eq:Seff_decomp_def}
\end{eqnarray}
where $\nu=(b-4)/2$ is a coefficient 
with the Gell-Mann--Low coefficient $b$
given by Eq.~(\ref{eq:2loop}), and 
$\bar{\rho}^2$ is the average size squared 
of instantons in the full ensembles including 
the instanton-instanton and instanton-quark 
interactions. The variational single 
instanton distribution is used as $Z(\alpha=0) \equiv Z_0$. 
Its form is given by
\begin{eqnarray}
  Z_0 
  &=& \frac{1}{N_+! N_-!} \(V\mu_0\)^{N_+ + N_-},
  \label{eq:Z0}
\end{eqnarray}
with 
\begin{eqnarray}
  \mu_0 
  &=& \int_0^\infty d\rho~ f(\rho) \exp \( -\nu \frac{\rho^2}{\bar{\rho}^2} \).
  \label{eq:Variational_ansatz}
\end{eqnarray}

\subsection  {Quark condensate in instanton ensemble}
\label{subsec:qq}

  In the instanton liquid model, 
the quark condensate is evaluated as 
the expectation value of the traced 
quark propagator at the same space-time coordinate as follows:
\begin{eqnarray}
  \braket{\bar{q}q} 
  &=& \sum_{A,\alpha} \braket{q^\dag(x)^A_\alpha q(x)^A_\alpha} \nonumber \\
  &=& - \lim_{y \to x} \sum_{A,\alpha} \braket{q(x)^A_\alpha q^\dag(y)^A_\alpha}  \nonumber \\
  &=& - \lim_{y \to x} \frac{1}{Z} \int D \Omega~\Tr \[S(x,y;m)\] e^{-S_{\rm int}} \Det (\slh{D}+m). \nonumber \\
\end{eqnarray}
Here we write the measure of the path integral 
as $D\Omega$ in short that is
given in the partition function (\ref{eq:IILM}), 
$A=1,\dots,N_c$ and $\alpha=1,\dots,4$ represent the color 
and the Dirac indices, respectively and $\Tr$ taken for 
the both indices.
The quark propagator is approximated as a sum of 
contributions from the free and the zero-mode propagators 
by inverting the Dirac operator in the basis spanned by 
the quark zero-mode wave functions in instantons background
as following \cite{Schafer:1998}
\begin{eqnarray}
  &\ & S(x,y;m) 
  \approx S_0(x,y) + S^{\rm ZM}(x,y;m), \nonumber \\
  &\ & S_0(x,y)=\frac{i}{2\pi^2} \frac{\g \cdot (x-y)}{(x-y)^4}, \nonumber \\
  &\ & S^{\rm ZM}(x,y;m) =
  \sum_{I,J}
  \[
  \psi_{0,I}(x)
  \[(-iT + m)^{-1}\]_{IJ}
  \psi^{\dagger}_{0,J}(y)
  \], \nonumber \\ 
  \label{eq:S_full}
\end{eqnarray}
where the matrix $T$ is the overlap matrix 
given in Eq.~(\ref{eq:matrix_T}). Here we omit to write
the collective coordinates of instantons $\{ U,\rho,z\}$
from the argument of the quark zero-mode wave functions.
We obtain the quark condensate by averaging it 
over the configurations.

\subsection  {Monte Carlo simulation}
\label{subsec:MC_simulation}

  The simplest way to simulate the instanton liquid 
model described by the Euclid partition function 
(\ref{eq:IILM}) is to use the Monte Carlo method with 
a weight function $S_{\rm eff}$ given by 
\begin{eqnarray}
  S_{\rm eff} = 
  - \sum_{i=1}^{N_+ + N_-} \ln [f(\rho_i)] 
  + S_{\rm int} 
  - \sum_{f=1}^{N_f} \ln \Det (\slh{D} + m_f).
  \nonumber\\ \label{eq:Seff}
\end{eqnarray}
To perform Monte Carlo simulations using 
the Markov Chain Monte Carlo (MCMC) method, 
it is crucial to understand the weight function $S_{\rm eff}$. 
This function represents the probability density $p(\Omega_i)$
as $p(\Omega_i) \propto \exp [-S_{\rm eff}(\Omega_i)],$ 
where $\Omega_i$ denotes a configuration within the 
considered ensemble. Once the partition function is established, 
we derive the weight function, as illustrated in Eq.~(\ref{eq:Seff}).
Employing algorithms such as the Metropolis algorithm or Hybrid Monte Carlo (HMC) 
algorithm, we generate a series of configurations 
$\{\Omega\} = (\Omega_1, \Omega_2,\dots, \Omega_{N_{\rm conf}})$ 
that converge to the given probability density function because of 
the detailed balance condition of the algorithms. Here, $N_{\rm conf}$ 
represents the number of configurations.

  The expectation value $\braket{O}$ of an operator $O$ 
that is expressed as a function of a configuration $\Omega_i$ 
is computed from these configurations using the formula:
\begin{eqnarray}
  \braket{O} = \lim_{N_{\rm conf} \to \infty} 
  \frac{1}{N_{\rm conf}} \sum_{i=1}^{N_{\rm conf}}
  O(\Omega_i). \label{eq:Exp_value}
\end{eqnarray}
For more details on the Monte Carlo 
simulations using the MCMC, we referred 
to some textbooks and an introductory article 
\cite{Binder:2010,Randau:2014,Hanada:2018}.

\section  {Results}
\label{sec:Results}

  In this section, we will show our numerical results. 
In subsection \ref{subsec:Computational_setup}, 
we explain the computational setup and numerical method
used for our calculations. 
In subsection \ref{subsec:F_res}, 
we show our results of the free energy as a function 
of the instanton density. 
In subsection \ref{subsec:qq_res}, 
we present the results of the quark condensate 
as a function of the instanton density. 
Combining these results, in subsection \ref{subsec:F_vs_qq}, 
we obtain the free energies as a function of the quark condensate 
and analyze them near the origin.

\subsection  {Computational setup}
\label{subsec:Computational_setup}

  In our simulations, each configuration $\Omega_i$ consists of 
$N_+$ instantons and $N_-$ anti-instantons labeled by 
their collective coordinate $\{U,\rho,z\}$ and 
the partition function for each instanton density 
is determined by generating 5000 configurations 
after 1000 initial sweeps with $N=N_++N_-=16+16$ instantons 
and anti-instantons. 
The simulations with different instanton density $n=N/V$ are achieved by 
changing the simulation box size $V = L^4$
with the fixed number of instantons $N$. 
Whole simulations are performed 
under the periodic boundary condition 
for the coordinates of instantons and 
anti-instantons.
All quantities in this calculation are 
nondimensionalized by $\Lambda$ which 
has been introduced through the $\beta$-functions (\ref{eq:1loop}).
The value of the scale parameter $\Lambda$ 
is determined so that the free energy 
has a minimum value at the instanton 
density $n=1~{\rm fm^{-4}}$ \cite{Schafer:1996}.
This density with the minimum free energy 
is the vacuum instanton density.
We are interested in the quark condensate dependence 
of the free energy in the small quark mass regime, 
so we set the current quark masses $m$ to be as small as 
possible within a stable run of the simulations.
The calculations below are performed with 
the current quark masses 
$m=0.1\Lambda, 0.15\Lambda, 0.2\Lambda$.

  In Table~\ref{table:Quench_Unquench_F_result}, 
we show the bulk parameters, 
such as the current quark mass $m$, 
the instanton density in the vacuum $n$, 
the scale parameter $\Lambda$
and the instanton average size $\bar{\rho}$ 
in our simulations.
These values are consistent 
with the previous work \cite{Schafer:1996}. 
By fixing the unit such that $n=1~{\rm fm}^{-4}$ at the vacuum,
the values of the scale parameter are determined as $\Lambda = 373,360,351~{\rm MeV}$ 
for each current quark mass. 
Using these values, the average instanton sizes 
in our calculations are evaluated as $\bar{\rho}=0.35$ to $0.37~{\rm fm}$ 
compared to $\bar{\rho}=0.42~{\rm fm}$ in Ref.~\cite{Schafer:1996}.

  To calculate the free energy by Eq.~(\ref{eq:F_def}), 
we need the value of the partition function.
The partition function is obtained 
through Eq.~(\ref{eq:Z_alpha_cal}),
so we first calculate $Z_0$ with 
the variational single instanton distribution (\ref{eq:Variational_ansatz}).
The single instanton distribution can be evaluated 
from initial 1000 sweeps with full interaction $\alpha=1$. 
We calculate the average size squared of instantons $\bar{\rho}^2$ 
appearing in Eq.~(\ref{eq:Variational_ansatz}) using 
the initial sweeps. The $\rho$-integral over an infinite 
interval $[0,\infty]$ appearing in Eq.~(\ref{eq:Variational_ansatz}) 
is practically performed over a finite interval $[0,\Lambda^{-1}]$.
The remaining task for the calculation 
of the partition function is to perform 
the integral by summing the integrands.
This integral is done at 10 different coupling 
values $\alpha$.

  In the computation of the quark condensate, 
only the quark zero-mode propagator 
$S^{\rm ZM}(x,y;m)$ in Eq.~(\ref{eq:S_full}) 
is evaluated to subtract the infinite contribution 
initiated by the free propagator $S_0(x,x)$ 
at the same space-time coordinate.
In the actual calculations, for each instanton density, 
the trace of the quark zero-mode propagator is averaged 
over the 5000 configurations, and also averaged 
over 10,000 different space-time coordinates 
to reduce the effect of incomplete equilibration 
of the configurations. 

\begin{table}
  \caption{
    \label{table:Quench_Unquench_F_result}
  The bulk parameters at the vacuum instanton 
  density. The values of the current quark mass $m$, 
  the vacuum instanton density $n$, 
  the scale parameter $\Lambda$ and 
  the instanton average size $\bar{\rho}$ 
  are given. 
  All quantities are given in both the physical unit and $\Lambda$. 
  The numbers in square brackets represent 
  the unit of $\Lambda$. 
  The superscript $^*$ represents the input parameter.
  }
  \begin{tabular*}{86mm}{lcrcrcr}
    \hline \hline
    $m~({\rm MeV})$         & \hspace{1.3em}  & $37.3~[0.1^*]$    & \hspace{1.3em} & $54.1~[0.15^*]$   & \hspace{1.3em} & $70.2~[0.2^*]$   \\ \hline 
    $n~({\rm fm^{-4}})$     &               & $1.00^*~[0.079]$  &              & $1.00^*~[0.090]$  &              & $1.00^*~[0.10]$  \\ 
    $\Lambda~({\rm MeV})$   &               & $373~[1]$         &              & $360~[1]$        &              & $351~[1]$        \\ 
    $\bar{\rho}~({\rm fm})$ &               & $0.35~[0.66]$     &              & $0.36~[0.66]$    &              & $0.37~[0.66]$    \\ 
    \hline \hline 
  \end{tabular*}
\end{table}

\subsection  {Free energy}
\label{subsec:F_res}

\begin{figure}
  \includegraphics[width=86mm]{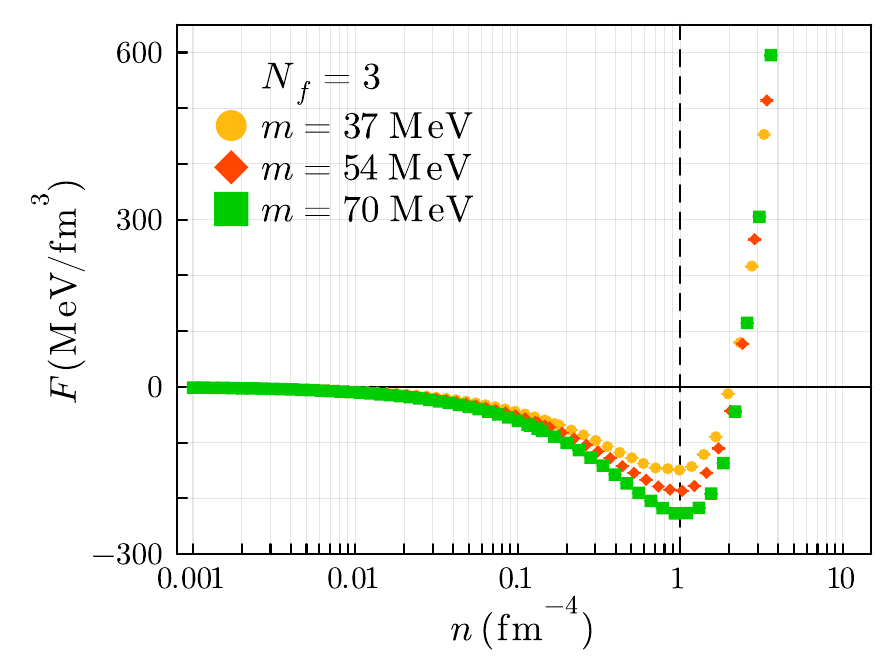}
  \caption{
    \label{fig:F_vs_n_Quench_Unquench}
    The free energies as a function of the instanton density 
    for different current quark masses.
  }
\end{figure}

  In Fig.~\ref{fig:F_vs_n_Quench_Unquench}, 
we show our numerical results of 
the free energies (in the unit of ${\rm MeV/fm^3}$) 
versus the instanton density (in the unit of ${\rm fm^{-4}}$)
for three values of the current quark mass.
Our results are consistent with the previous work \cite{Schafer:1996}.
The free energy is monotonically decreasing 
towards a minimum value and then rapidly 
increasing at higher instanton density. 
This shows that attractive interaction is 
dominant in lower instanton density regions, while
repulsive interaction becomes important 
at higher instanton density.

\begin{table}
  \caption{
    \label{table:Quench_Unquench_qq_result}
  The free energy and the quark condensate 
  at the vacuum instanton density with different 
  current quark masses.
  The notation is the same 
  as in Table~\ref{table:Quench_Unquench_F_result}. 
  }
  \begin{tabular*}{86mm}{rrrr} \hline \hline 
    $m({\rm MeV})$                          & $37~[0.1^*]$      & $54~[0.15^*]$     & $70~[0.2^*]$         \\ \hline
    $F({\rm MeV/fm^3})$                     & $-149[-0.060]$    & $-187[-0.085]$    & $-228[-0.116]$       \\   
    $-\braket{\bar{q}q}^{1/3}({\rm MeV})$ & $-188[-0.12]$     & $-196[-0.16]$     & $-200[-0.18]$        \\ \hline \hline 
  \end{tabular*}
\end{table}

  In Table~\ref{table:Quench_Unquench_qq_result}, 
we summarize our numerical results of 
the free energy and the quark condensate 
at the vacuum for three different current quark 
masses $m=37, 54, 70~{\rm MeV}$. 
By using the values of the scale parameter 
in Table~\ref{table:Quench_Unquench_F_result},
the free energies are evaluated as 
$F = -149,-187,-228~{\rm MeV/fm^3}$. 
For reference, we perform further calculations 
for the current quark masses with values close 
to those in the previous work, 
such as $m=96~{\rm MeV}$ \cite{Schafer:1996}.
That shows a good agreement with the previous work, 
and we conclude that our simulations work well.

\subsection  {Quark condensate}
\label{subsec:qq_res}

\begin{figure}
  \includegraphics[width=86mm]{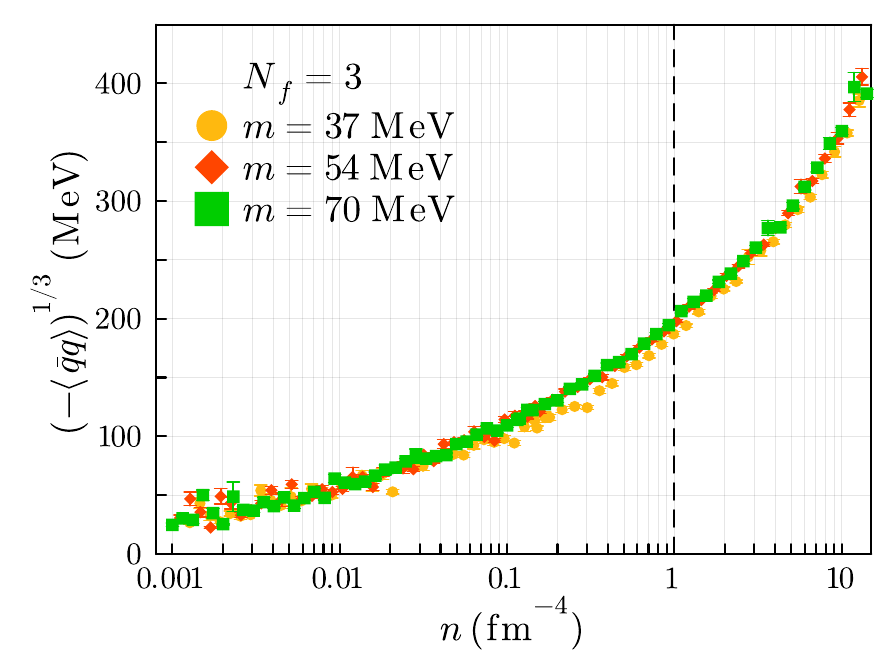}
  \caption{
    \label{fig:qqn_Quench_Unquench}
  The quark condensates as a function of the instanton density.
  Details about the different quark masses are explained in the main text.
  }
\end{figure}

  In Fig.~\ref{fig:qqn_Quench_Unquench}, 
we show the instanton density dependence of 
cubic root of the quark condensate $(-\braket{\bar{q}q})^{1/3}$ 
in the unit of ${\rm MeV}$.
We find that the absolute values 
of the quark condensate
increase monotonically 
as the instanton density increases.
We also find that the value of the quark 
condensates at the vacuum is insensitive 
to the value of the current quark masses.
This means that the contribution from 
the explicit breaking of chiral 
symmetry due to the current quark mass
is not so large for the value of the 
quark condensate.

  In Table~\ref{table:Quench_Unquench_qq_result}, 
we show the values of the quark condensate 
at the vacuum instanton density.
Our results almost reproduce 
the empirical values obtained 
by various previous works 
\cite{GellMann:1968,Reinders:1985,Dosch:1998,Harnett:2021,Giusti:2001,McNeile:2013,Borsanyi:2013,Durr:2014,Cossu:2016,Davies:2019,FLAG:2021,Gasser:1985,Jamin:2001,Boyle:2016,Kneur:2020}. 
We obtain the values of the quark condensate as
$|\braket{\bar{q}q}|^{1/3} = 188, 196, 200~{\rm MeV}$
at the vacuum with the current quark masses 
$m=37,54,70~{\rm MeV}$, respectively.
These values give a good estimate of 
the quark condensates although they are slightly 
smaller in the absolute value.

\subsection  {Free energy vs.~quark condensate}
\label{subsec:F_vs_qq}

\begin{figure}
  \includegraphics[width=86mm]{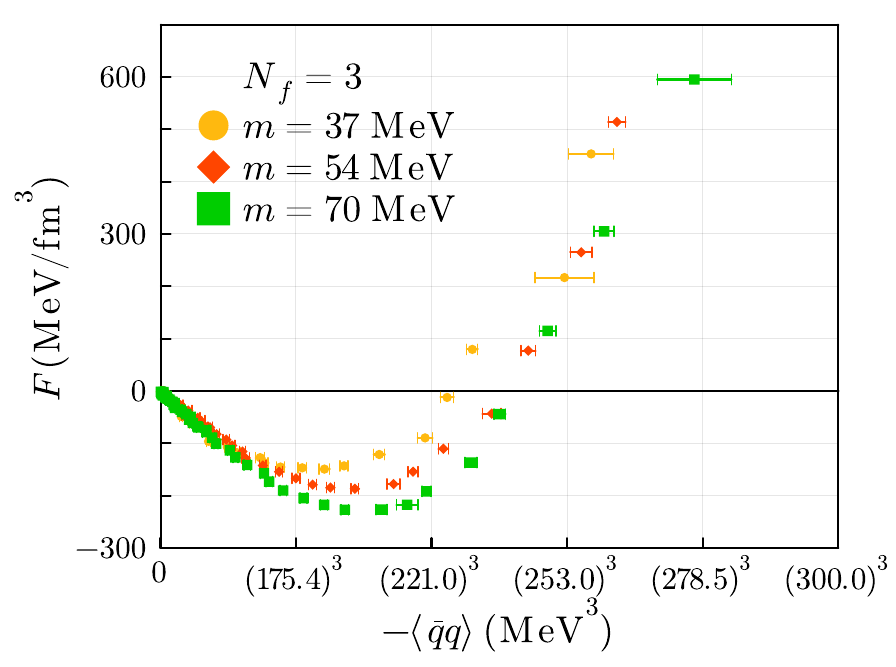}
  \caption{
    \label{fig:F_vs_qq_Quench_Unquench}
  The free energy densities as a function of the quark condensate
  for three quark masses. The free energy and the quark 
  condensate are given in the unit of ${\rm MeV/fm^3}$ 
  and ${\rm MeV^3}$, respectively.
  }
\end{figure}

  Combining the results of the free energy 
and the quark condensate, we obtain the quark 
condensate dependence of the free energy 
as shown in Fig.~\ref{fig:F_vs_qq_Quench_Unquench}. 
The free energies monotonically decrease 
towards the minimum values as the quark 
condensate increases in magnitude.
Once the free energies have the minimum value, 
they start to increase as expected from 
the instanton density dependence of them 
(also see Fig.~\ref{fig:F_vs_n_Quench_Unquench}). 
From this, we observe that the free energy has 
its minimum value at the point with the finite 
value of the quark condensate. This shows 
that chiral symmetry is dynamically broken 
in the vacuum of the IILM. 

  The behavior of the free energy near the origin
appears to be decreasing in a downward convex trend.
This trend is crucial for the sign 
of the curvature of the free energy 
at the origin. In other words, it provides one of the hints 
for discriminating patterns of chiral symmetry breaking 
through our definition of the determination procedure 
for the anomaly-driven breaking discussed in Sect.~\ref{subsec:ADbreaking_general}.
So we study the quark condensate dependence 
of the free energy more quantitatively.

  We aim to evaluate the curvature of 
the free energy at the origin from our simulation results of the IILM. 
For that, we perform polynomial fits for our three data 
sets with different current quark masses.
Each data set consists of $N_{\rm data}$ pairs 
of $\braket{\bar{q}q}$ and $F$, 
i.e., $(|\braket{\bar{q}q}|_i, F_i), i=1,\dots,N_{\rm data}$ 
arranged in ascending order, where 
$N_{\rm data}$ is the number of data of the free energy and 
the quark condensate and it is equal to 
the number of the grid points of the instanton density, 
$N_{\rm data} = 71$ in our calculations.

  We use a polynomial function given by 
\begin{eqnarray}
  F(\braket{\bar{q}q}) = \sum_{j=0}^{K} C_j \braket{\bar{q}q}^j,
  \label{eq:fitmodel_general}
\end{eqnarray}
for the fitting model in this analysis.
We perform the fits 
for four orders $K=1,\dots,4$.
For each order $K$, 
we optimize the parameters $\{C_j\}~(j=0,\dots,K)$ 
so that they minimize the reduced 
chi-square including errors of 
both axes as given in Ref.~\cite{Orear:1984} by 
\begin{eqnarray}
  \chi^2_{\rm d.o.f.}
  = \frac{1}{N_{\rm d.o.f.}} \sum_{i=1}^{M} \frac{(y_i - f(x_i))^2}{\sigma_{y_i}^2 + \sigma_{x_i}^2 \[f'(x_i)\]^2},
  \label{eq:chi_square}
\end{eqnarray}
where $(x_i,y_i)$ with their
errors $(\sigma_{x_i}, \sigma_{y_i})$ 
correspond to our data 
set $(|\braket{\bar{q}q}|_i,F_i)$ 
with their errors, $f(x)$ and $f'(x)$ represent 
the fit model (\ref{eq:fitmodel_general})
and its first derivative, $N_{\rm d.o.f.}$ is 
the number of degrees of freedom which is defined by 
$N_{\rm d.o.f.}\equiv M-(K+1)$, and $M$ is the number of 
data used in the fit. 
We use the data from $i=1$ to $i=M$ 
rather all data because we want to 
know the behavior of the free energy 
around the origin.

  We determine an upper 
limit of fit range, $M$, as follows.
We calculate the reduced chi-square 
(\ref{eq:chi_square}) as we increase 
the number of data that we use and obtain 
the reduced chi-square as a function 
of the number of data. We then find 
the number of data for which the reduced 
chi-square has a local minimum value. 
We finally use this number of data 
as the value of $M$ for fitting. 
Here we skip a trivial local 
minimum obtained with a small number of data.
The specific values of $M$ used in each fit, 
the corresponding quark condensate values and the reduced 
chi-square are summarized in Table~\ref{table:M_and_qq_results_for_fit}.

\begin{table}
  \caption{
    Values of $M$, the cubic root of the absolute value of 
    the quark condensate in the unit of ${\rm MeV}$
    and the reduced chi-square for each fit with the current quark mass $m$. 
    The values of current quark mass are given in the unit of ${\rm MeV}$.
  }
  \label{table:M_and_qq_results_for_fit}
  \begin{tabular*}{86mm}{ccccccc} \hline \hline 
    $m$ & \hspace{1.5em} & 37 & \hspace{1.5em} & 54 & \hspace{1.5em} & 70 \\ \hline 
    $K$ & & \multicolumn{5}{c}{$(M, |\braket{\bar{q}q}|^{1/3}, \chi^2_{\rm d.o.f.})$} \\
    1 & & $(41, 125, 3.87)$ & & $(25,  79, 3.84)$ & & $(40, 131, 4.22)$ \\
    2 & & $(39, 123, 3.88)$ & & $(48, 189, 3.61)$ & & $(48, 187, 3.48)$ \\
    3 & & $(47, 178, 4.01)$ & & $(56, 263, 3.16)$ & & $(58, 278, 3.06)$ \\
    4 & & $(48, 187, 4.01)$ & & $(62, 353, 3.05)$ & & $(58, 278, 3.11)$ \\ 
    \hline \hline
  \end{tabular*}
\end{table}

\begin{table}
  \caption{
    Fit results of $C_0$ for each order $K$ and current quark mass $m$. 
    Current quark mass $m$ is given in the unit of ${\rm MeV}$ 
    and the coefficient $C_0$ is given in the unit of ${\rm MeV/fm^3}$.
  }
  \label{table:C0_results}
  \begin{tabular*}{86mm}{ccccccc} \hline \hline
    $m$ & \hspace{2.3em} & 37 & \hspace{2.3em} & 54 & \hspace{2.3em} & 70 \\ \hline
    $K$ & & \multicolumn{5}{c}{$C_0({\rm MeV/fm^3})$} \\ 
    1 & & $-0.64^{+0.07}_{-0.07}$ & & $-1.47^{+0.04}_{-0.04}$ & & $-1.51^{+0.06}_{-0.06}$ \\ 
    2 & & $-0.51^{+0.08}_{-0.08}$ & & $-1.48^{+0.04}_{-0.04}$ & & $-1.44^{+0.06}_{-0.06}$ \\
    3 & & $-0.47^{+0.08}_{-0.08}$ & & $-1.51^{+0.04}_{-0.04}$ & & $-1.48^{+0.06}_{-0.06}$ \\
    4 & & $-0.52^{+0.07}_{-0.07}$ & & $-1.52^{+0.04}_{-0.04}$ & & $-1.47^{+0.06}_{-0.06}$ \\
    \hline \hline 
  \end{tabular*}
\end{table}

  In Table~\ref{table:C0_results},
we show the fit results of $C_0$.
We find that the value of 
the coefficient $C_0$ is close to zero.
For different fit orders and 
quark masses, 
the values of $C_0$ range
from $-1.52$ to $-0.47$
in the unit of ${\rm MeV/fm^3}$.
These values are sufficiently 
small compared to the typical value of 
the free energy $F$ and 
we conclude that the coefficient $C_0$ 
is consistent with zero.

\begin{table}
  \caption{
    Fit results of $C_1$ for each order $K$ and current quark mass $m$.
    The coefficient $C_1$ is given in the unit of ${\rm MeV}$.
  }
  \label{table:C1_results}
    \begin{tabular*}{86mm}{ccccccc} \hline \hline
      $m$ & \hspace{3em} & 37 & \hspace{3em} & 54 & \hspace{3em} & 70 \\ \hline
      $K$ & & \multicolumn{5}{c}{$C_1({\rm MeV})$} \\ 
      1 & & $364.0^{+5.6}_{-5.6}$ & & $332.5^{+9.1}_{-9.7}$ & & $349.3^{+4.5}_{-4.6}$ \\ 
      2 & & $395.8^{+5.7}_{-6.0}$ & & $326.2^{+1.4}_{-1.4}$ & & $371.8^{+1.7}_{-1.8}$ \\
      3 & & $403.9^{+0.6}_{-0.6}$ & & $312.6^{+0.4}_{-0.4}$ & & $358.1^{+0.7}_{-0.7}$ \\
      4 & & $391.9^{+0.6}_{-0.6}$ & & $307.9^{+0.6}_{-0.4}$ & & $361.8^{+0.7}_{-0.7}$ \\
      \hline \hline 
    \end{tabular*}
\end{table}

  In Table~\ref{table:C1_results},
we show the fit results of $C_1$. 
We find that the value of $C_1$ is insensitive 
to the fitting order $K$. This implies that 
the value of $C_1$ is well determined 
around the origin. We find also that $C_1$ 
has no clear quark mass dependence. 
The averaged values of $C_1$ 
over the different orders are 
$389~{\rm MeV},320~{\rm MeV}$ and $360~{\rm MeV}$ 
for $m=37,54$ and $70~{\rm MeV}$.

\begin{table}
  \caption{
    Fit results of $C_2$ for each order $K$ and current quark mass $m$.
    The coefficient $C_2$ is given in the unit of ${\rm MeV^{-2}}$.
  }
  \label{table:C2_results}
    \begin{tabular*}{86mm}{ccccccc} \hline \hline 
      $m$ & \hspace{3em} & 37 & \hspace{3em} & 54 & \hspace{3em} & 70 \\ \hline
      $K$ & & \multicolumn{5}{c}{$C_2 \times 10^5({\rm MeV^{-2}})$} \\ 
      1 & & - & & - & & -  \\ 
      2 & & $3.46^{+0.52}_{-0.58}$ & & $1.71^{+0.03}_{-0.03}$ & & $1.74^{+0.03}_{-0.04}$ \\
      3 & & $3.51^{+0.01}_{-0.01}$ & & $0.78^{+0.01}_{-0.01}$ & & $0.79^{+0.01}_{-0.01}$ \\
      4 & & $1.84^{+0.01}_{-0.01}$ & & $0.44^{+0.01}_{-0.01}$ & & $0.99^{+0.01}_{-0.01}$ \\
      \hline \hline 
    \end{tabular*}
\end{table}

  In Table~\ref{table:C2_results},
we show the fit results of $C_2$.
We find that the coefficients $C_2$ 
appear to be positive. For each quark mass, 
the averaged values of $C_2$ 
over the different fit orders 
are $C_2=(2.94,0.98,1.17)\times 10^{-5}~{\rm MeV^{-2}}$ 
for $m=37,54$ and $70~{\rm MeV}$.
These values of $C_2$ are located 
at the positive region for all orders 
and current quark masses. This suggests 
that the curvature of the free energy 
at the origin is positive for the IILM.

\begin{figure}
  \includegraphics[width=84mm]{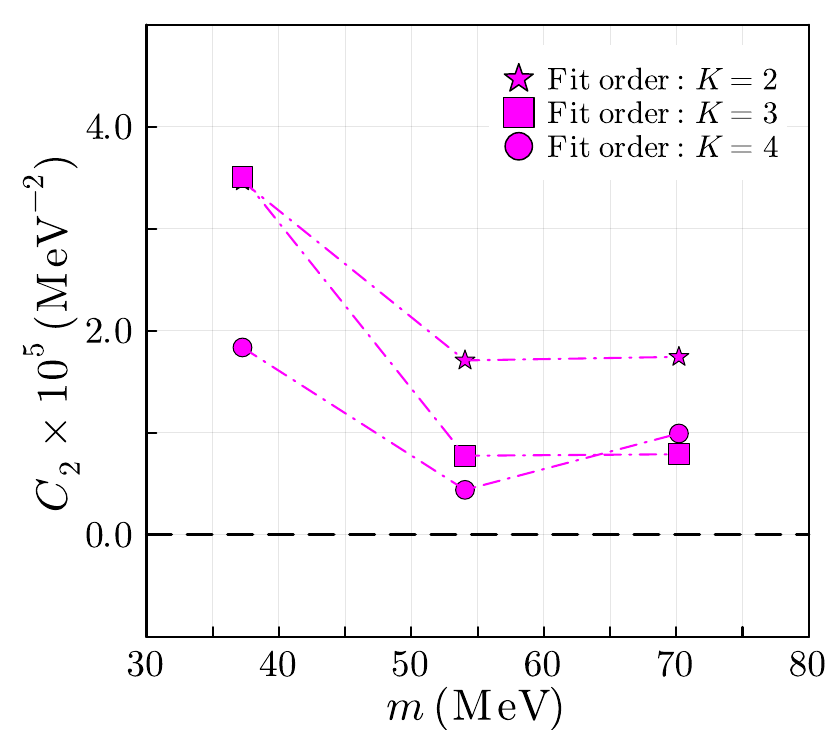}
  \caption{
    \label{fig:m_vs_C2_analysis} 
    Current quark mass dependence of the coefficient $C_2$. 
    Fit results of different orders are shown 
    with different types of markers.
    Error bars are omitted because they are small.
    }
\end{figure}

  In Fig.~\ref{fig:m_vs_C2_analysis}, 
we show the current quark mass dependence 
of the coefficient $C_2$. 
The value of $C_2$ is not very sensitive 
to the value of the current quark mass.
For the fit with the smallest quark mass, 
the values of $C_2$ are evaluated as slightly 
larger than the results of other two quark masses. 
These results suggest that the value of $C_2$ and 
the curvature of the free energy at the origin
is positive for wide current quark mass region
and thus we conclude that the anomaly-driven breaking 
of chiral symmetry is realized in the IILM 
by our definition (\ref{eq:Condition_ADB})

  Here we do not show the values of 
the coefficients $C_3$ and $C_4$
because these coefficients are introduced 
to confirm the stability of fits
of $C_0, C_1$ and $C_2$ with and
without these higher-order terms. 
Furthermore, since we use the part 
of full data near the origin 
for fits, the values of $C_3$ and $C_4$ 
are not determined well and irrelevant 
to our discussion.

  Interestingly we find the opposite sign of $C_2$ 
in the quenched calculations where 
the quark determinant part in 
the partition function (\ref{eq:IILM}) is set as
$\prod_f \Det (\slh{D}+m_f)=1$
in generating the configurations. 
The details of the quenched calculation are shown in Appendix~\ref{App:1}.
As we show in Table \ref{table:Quench_C2_results}, 
the quenched calculations 
suggest the negative values of coefficient $C_2$.
By our definition of the anomaly-driven breaking (\ref{eq:Condition_ADB}), 
this concludes that the normal breaking is realized 
in the quenched IILM. The opposite signs of $C_2$ 
obtained by the full and quenched calculations 
mean that the different scenarios of chiral symmetry breaking 
are possible depending on the presence of the quark determinant part 
in the IILM partition function. Since the quark determinant 
contributes to the instanton-quark interaction 
in the ensemble, this implies that the quarks play a 
crucial role in the anomaly-driven breaking for the IILM. 

\begin{table}
  \caption{
    Fit results of $C_2$ in the quenched calculations
    for each order $K$ and current quark mass $m$.
    The coefficient $C_2$ is given in the unit of ${\rm MeV^{-2}}$.
  }
  \label{table:Quench_C2_results}
    \begin{tabular*}{86mm}{ccccccc} \hline \hline 
      \multicolumn{7}{l}{Quenched} \\ 
      $m$ & \hspace{2.5em} & 2.8 & \hspace{3em} & 14 & \hspace{3em} & 28 \\\hline
      $K$ & & \multicolumn{5}{c}{$C_2 \times 10^5 ({\rm MeV^{-2}})$} \\ 
      1 & & - & & - & & - \\
      2 & & $-3.97^{+0.08}_{-0.08}$ & & $-3.48^{+0.11}_{-0.11}$ & & $-2.40^{+0.11}_{-0.11}$ \\
      3 & & $-3.32^{+0.10}_{-0.10}$ & & $-4.39^{+0.07}_{-0.07}$ & & $-3.22^{+0.08}_{-0.08}$ \\
      4 & & $-2.72^{+0.26}_{-0.23}$ & & $-3.61^{+0.01}_{-0.01}$ & & $-2.11^{+0.01}_{-0.01}$ \\
      \hline \hline 
    \end{tabular*}
\end{table}

\section  {Conclusions}
\label{sec:Conclusions}

  We have examined the possibility of 
the chiral symmetry breaking scenario 
that can be realized when the \UA anomaly 
contribution is sufficiently large using the IILM.
Following the NJL model,
we have generalized the determination procedure for 
the pattern of chiral symmetry breaking.
We use the second-order differential coefficient
of the vacuum energy density with respect to 
the quark condensate at the origin, e.g., the curvature.
We have defined that the pattern of 
chiral symmetry breaking is the normal or 
the anomaly-driven one if the curvature is 
negative or positive, respectively. 

  We have found that the curvature 
appears to be positive in the IILM. 
This means that the anomaly-driven breaking 
is feasible in the IILM by our definition.
In contrast to that, the quenched calculations 
show the negative curvature and support the realization 
of the normal chiral symmetry breaking 
in the quenched IILM. 
These results indicate that 
the instanton-quark interaction 
is crucial to realize the anomaly-driven breaking because 
the difference between the full and quenched IILM
is the presence or absence of the quark determinant 
that contributes to the interaction among instantons 
and the dynamical quarks.

  One important direction for future development is to 
investigate the mass of mesons, 
such as $\sigma$ and $\eta'$. 
As discussed in the previous work \cite{Kono:2021}, 
chiral effective theories concluded that the $\sigma$ mass 
could be smaller than $800~{\rm MeV}$ in the anomaly-driven 
symmetry breaking. We expect the same conclusions 
to be drawn in the IILM. Even if such conclusions 
cannot be reached, it is interesting to clarify 
the differences between the chiral effective theories 
and the IILM. 
Computation with different flavors, e.g., $N_f=2$, 
using the IILM may provide further insights into 
the anomaly-driven breaking in QCD.
These are subjects to be investigated in future studies.

%
%
\section*{Acknowledgments}
\noindent
We would like to thank Masayasu Harada for 
useful discussions and comments.
We also thank Masakiyo Kitazawa 
and Kotaro Murakami for their advice.
This work of Y.S. was partly supported by the 
Advanced Research Center for Quantum Physics and 
Nanoscience, Tokyo Institute of Technology, 
and JST SPRING, Grant Number JPMJSP2106. 
The work of D.J. was supported in part 
by Grants-in-Aid for Scientific Research from 
JSPS (JP21K03530, JP22H04917 and JP23K03427).
%
%
%

%
%
%

\appendix 
\section {Results of the quenched calculations}
\label{App:1}

  We show the numerical results for the quenched calculations.
The quenched calculations can be performed in the same way 
as in the full calculations except for setting 
the quark determinant in the partition function to unity.
The following numerical results of the quenched calculations 
are obtained by the same setup as the full calculations, 
that is, 5000 configurations after 1000 initial sweeps 
with $N=N_++N_-=16+16$ instantons and anti-instantons.
In the quenched calculations, 
the current quark mass enters only 
the calculation of the quark condensate 
through the quark zero-mode propagator (\ref{eq:S_full}).

  In Table~\ref{table:Quench_bulk_param}, we show 
the bulk parameters of the quenched ensemble. 
These values are consistent 
with the quenched calculation 
in the previous work \cite{Schafer:1996}. 
We obtain the scale parameter $\Lambda=281~{\rm MeV}$ and 
the average instanton size $\bar{\rho}=0.42$, 
which are compared to 
$\Lambda = 270~{\rm MeV}$ and $\bar{\rho} = 0.43~{\rm fm}$ 
in the previous work~\cite{Schafer:1996}, respectively.
Our result of the free energy
$F=-532~{\rm MeV/fm^3}$
shows a good agreement with 
$F=-526~{\rm MeV/fm^3}$ 
for the quenched calculation in Ref.~\cite{Schafer:1996}.
Our result also shows a good agreement with 
the estimation from the trace anomaly $F = -543~{\rm MeV/fm^4}$
as discussed in Ref.~\cite{Schafer:1996}.

\begin{table}
  \caption{
    \label{table:Quench_bulk_param}
  The bulk parameters at the vacuum instanton 
  density for the quenched calculation. 
  The asterisk represents 
  the value of the input fix parameter.
  }
  \begin{tabular*}{86mm}{clcr}
    \hline \hline
    Quenched\hspace{2em} &    & \hspace{3em} &                  \\ \hline
    & $n~({\rm fm^{-4}})$     &              & $1.00^*~[0.24]$  \\
    & $\Lambda~({\rm MeV})$   &              & $281~[1]$        \\
    & $\bar{\rho}~({\rm fm})$ &              & $0.42~[0.60]$    \\
    & $F({\rm MeV/fm^3})$     &              & $-532~[-0.657]$  \\
    \hline \hline 
  \end{tabular*}
\end{table}

\begin{table}
  \caption{
    \label{table:Quench_Fqq_result}
  The quark condensates at the vacuum for the quenched calculations
  with different quark masses. The asterisk represents 
  the value of the input fix parameter.
  }
  \begin{tabular*}{86mm}{rrrr}
    \hline \hline
    $m({\rm MeV})$                        & $2.8~[0.01^*]$    & $14~[0.05^*]$    & $28~[0.1^*]$  \\ \hline
    $-\braket{\bar{q}q}^{1/3}({\rm MeV})$ & $-246~[-0.67]$    & $-244[-0.65]$    & $-233[-0.57]$ \\ \hline \hline 
  \end{tabular*}
\end{table}

  In Table~\ref{table:Quench_Fqq_result}, 
we show the quark condensate at the vacuum
for three current quark masses.
Our results of the quark condensate 
$|\braket{\bar{q}q}|^{1/3}=246,244,233~{\rm MeV}$
for the current quark masses $m=2.8,14,28~{\rm MeV}$
almost reproduce the empirical values 
obtained by various previous works 
\cite{GellMann:1968,Reinders:1985,Dosch:1998,Harnett:2021,Giusti:2001,McNeile:2013,Borsanyi:2013,Durr:2014,Cossu:2016,Davies:2019,FLAG:2021,Gasser:1985,Jamin:2001,Boyle:2016,Kneur:2020}. 
We conclude that our simulations 
also work well in the quenched calculations.

\begin{figure}
  \includegraphics[width=86mm]{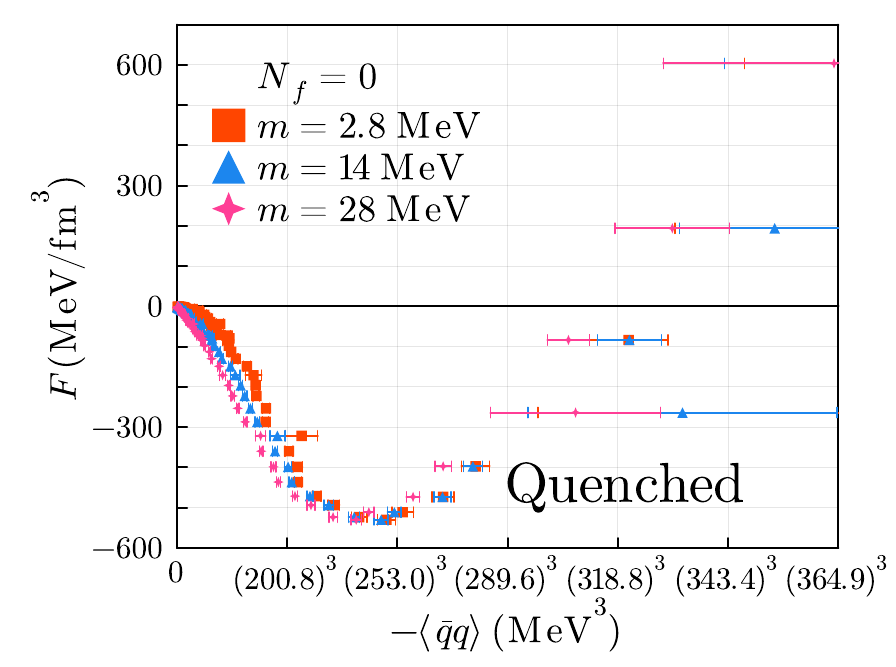}
  \caption{
    \label{fig:F_vs_qq_Quench}
  The free energies as a function of the quark condensate in the quenched 
  calculations.
  }
\end{figure}

  In Fig.~\ref{fig:F_vs_qq_Quench},
we show the free energy versus the quark condensate 
for the quenched calculations. This shows globally almost 
same behavior with the results of the unquenched calculations, 
but the trend of the free energy near the origin appears 
to be slightly different from the unquenched results.
We perform the same analysis to the unquenched calculations 
using the polynomial fitting and show the fit results following.

\begin{table}
  \caption{
    Fit results of $C_0$ in the quenched calculations
    for each order $K$ and current quark mass $m$. 
    Current quark mass $m$ is given in the unit of ${\rm MeV}$ 
    and the coefficient $C_0$ is given in the unit of ${\rm MeV/fm^3}$.
  }
  \label{table:Quench_C0_results}
  \begin{tabular*}{86mm}{ccccccc} \hline \hline
    \multicolumn{7}{l}{Quenched} \\ 
    $m$ & \hspace{2.5em} & 2.81 & \hspace{3em} & 14.04 & \hspace{3em} & 28.08 \\ \hline
    $K$ & & \multicolumn{5}{c}{$C_0({\rm MeV/fm^3})$} \\ 
    1 & & $-0.47^{+0.04}_{-0.04}$ & & $-0.72^{+0.03}_{-0.03}$ & & $-0.76^{+0.02}_{-0.02}$ \\
    2 & & $-0.74^{+0.03}_{-0.03}$ & & $-0.76^{+0.02}_{-0.02}$ & & $-0.80^{+0.02}_{-0.02}$ \\
    3 & & $-0.70^{+0.03}_{-0.03}$ & & $-0.78^{+0.02}_{-0.02}$ & & $-0.81^{+0.02}_{-0.02}$ \\
    4 & & $-0.69^{+0.03}_{-0.03}$ & & $-0.76^{+0.02}_{-0.02}$ & & $-0.80^{+0.02}_{-0.02}$ \\
    \hline \hline 
  \end{tabular*}
\end{table}

\begin{table}
  \caption{
    Fit results of $C_1$ in the quenched calculations
    for each order $K$ and current quark mass $m$.
    The coefficient $C_1$ is given in the unit of ${\rm MeV}$.
  }
  \label{table:Quench_C1_results}
    \begin{tabular*}{86mm}{ccccccc} \hline \hline
      \multicolumn{7}{l}{Quenched} \\ 
      $m$ & \hspace{2.5em} & 2.81 & \hspace{3em} & 14.04 & \hspace{3em} & 28.08 \\ \hline
      $K$ & & \multicolumn{5}{c}{$C_1({\rm MeV})$} \\ 
      1 & & $64.1^{+2.1}_{-2.2}$ & & $180.9^{+3.6}_{-3.8}$ & & $329.6^{+3.9}_{-3.9}$ \\
      2 & & $38.6^{+1.5}_{-1.7}$ & & $168.3^{+2.5}_{-2.7}$ & & $308.1^{+3.1}_{-3.1}$ \\
      3 & & $41.9^{+1.7}_{-1.7}$ & & $162.1^{+2.4}_{-2.4}$ & & $302.3^{+2.8}_{-2.8}$ \\
      4 & & $43.8^{+1.4}_{-1.4}$ & & $166.3^{+1.4}_{-1.4}$ & & $308.4^{+1.0}_{-0.8}$ \\
      \hline \hline 
    \end{tabular*}
\end{table}

  In Table~\ref{table:Quench_C0_results},
we show the fit results of $C_0$ 
for the quenched calculations.
We find that the value of the coefficient $C_0$ 
is also close to zero.
The value of $C_0$ averaged over different orders 
and quark masses is $C_0=-0.77~{\rm MeV/fm}^3$.
This value is sufficiently small compared 
to the typical value of 
the free energy $F$ and we conclude 
that the coefficient $C_0$ is also 
consistent with zero in the quenched calculation.

  In Table~\ref{table:Quench_C1_results},
we show the fit results of $C_1$ 
for the quenched calculations.
We find that the coefficient $C_1$ increases 
monotonically as the current quark mass increases.
For each quark mass, 
we average the values of $C_1$ 
over the different orders 
and obtain $C_1=48~{\rm MeV}$, $169~{\rm MeV}$ and $312~{\rm MeV}$
for $m=2.8,14$ and $28~{\rm MeV}$, respectively. 
We conclude that the value of $C_1$ 
increases monotonically as the quark mass increases
for the quenched calculations.

\begin{figure}
  \includegraphics[width=84mm]{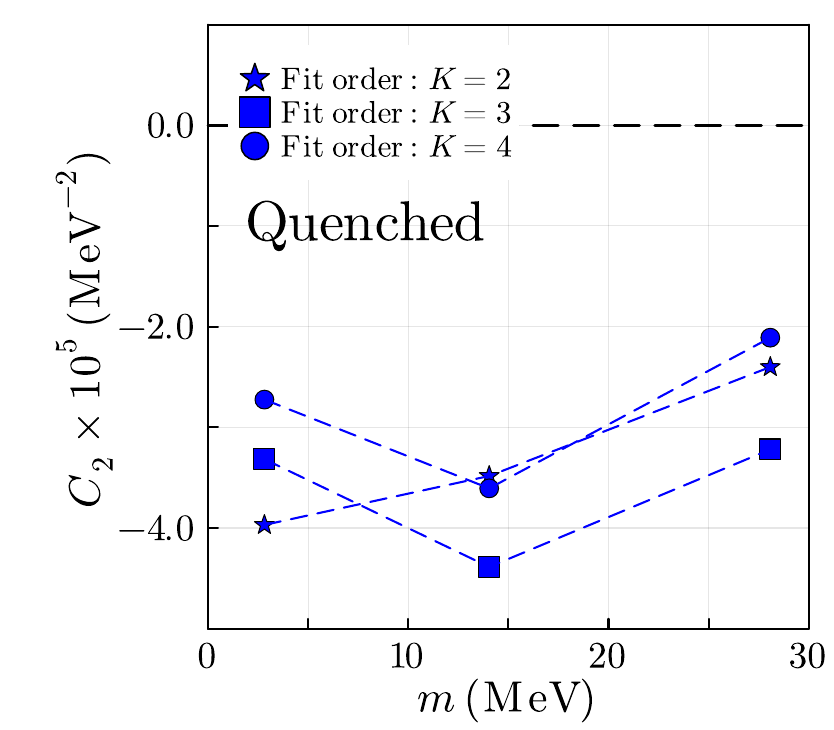}
  \caption{
    \label{fig:m_vs_C2_analysis_Quenched} 
    Current quark mass dependence of the coefficient $C_2$ 
    for the quenched calculations.
    Fit results of different orders 
    are shown with different markers.
    Error bars are omitted because they are small.
    Values are summarized in Table~\ref{table:Quench_C2_results}.
    }
\end{figure}

  In Fig.~\ref{fig:m_vs_C2_analysis_Quenched}, 
we show the current quark mass dependence of the 
coefficient $C_2$ for the quenched calculations.
For all quark masses, the value of $C_2$ appears 
to be negative in different fit orders. 
As we have discussed in Sect.~\ref{subsec:F_vs_qq}, 
these results suggest that 
the chiral symmetry is broken 
in the normal way for the quenched IILM
by our definition (\ref{eq:Condition_ADB}).

%
%

\end  {document}